\documentclass[manuscript]{aastex}

\usepackage{amsfonts}

\slugcomment{Not to appear in Nonlearned J., 45.}

\shorttitle{Energy dependence of QPO centroid frequency}
\shortauthors{QU et al.}

\begin{document}

\title{The energy dependence of the centroid frequency and phase lag
of the QPOs in GRS~1915+105}
\author{J. L. Qu\altaffilmark{1}, F. J. Lu\altaffilmark{1},
Y. Lu\altaffilmark{1}, L. M. Song\altaffilmark{1}, S.
Zhang\altaffilmark{1}, G. Q. Ding\altaffilmark{2}} \affil{Key
Laboratory for Particle Astrophysics, Institute of High Energy
Physics, CAS, Beijing, P. R. China}

\altaffiltext{1}{Key Laboratory for Particle Astrophysics, Institute
of High Energy Physics, Chinese Academy of Sciences (CAS), 19B
Yuquan Road, Beijing 100049, P. R. China; Email: qujl@ihep.ac.cn.}
\altaffiltext{2}{Urumqi Observatory, NAOC, 40-5 South Beijing Road,
Urumqi, Xinjiang 830011, P. R. China; Email: dinggq@uao.ac.cn}

\begin{abstract}
We present a study of the centroid frequencies and phase lags of the
quasi-periodic oscillations (QPOs) as functions of photon energy for
GRS~1915+105. It is found that the centroid frequencies of the
0.5-10 Hz QPOs and their phase lags are both energy dependent, and
there exists an anti-correlation between the QPO frequency and phase
lag. These new results challenge the popular QPO models, because
none of them can fully explain the observed properties. We suggest
that the observed QPO phase lags are partially due to the variation
of the QPO frequency with energy, especially for those with
frequency higher than 3.5 Hz.
\end{abstract}


\section{Introduction}

GRS~1915+105 is a low-mass X-ray black hole binary showing a rich
diversity of X-ray lightcurve morphology and complex timing
phenomena (Morgan et al. 1997; Cui 1999; Belloni et al. 2000; Ji et
al. 2003). The variability of the source can be reduced to
transitions between three basic states: a hard state corresponding
to the non-observability of the innermost part of the accretion disk
(state C), and two softer states with a fully observable disk but
different temperatures (states B and A) (Belloni et al. 1997a,b;
2000). According to the appearance of light curves and color-color
diagrams, the behaviors of the source can be further classified into
12 classes (Belloni et al. 2000). In additional to the above flux
and spectral variabilities, abundant quasi-periodic oscillations
(QPOs) are also observed in this system. The fundamental frequency
of its QPO ranges from mHz to several hundred Hz, and some QPOs are
detected up to the third harmonic (Morgan et al. 1997; Cui 1999).
According to their (fundamental) frequencies, the QPOs of
GRS~1915+105 can be divided into three classes: the low-frequency
($\sim$1-67 mHz) QPOs, the intermediate-frequency (0.5-10 Hz) QPOs,
and the high-frequency ($\sim$ 67 Hz) QPOs. These QPOs occur in
different states of the accretion disk of the source (Chakrabarti \&
Manickam 2000). The low-frequency QPOs and the high-frequency QPOs
are often observed during the soft state of GRS~1915+105 (state B).
The low-frequency QPOs are considered to be connected with disk
instability, as the rapid disappearance and refill of the inner
accretion disk (Belloni et al. 2000). The high-frequency QPO has
been proposed to arise in the close vicinity of the black hole and
been thought to reflect the general relativity properties of the
black hole (Cui et al. 1998). Conversely, the intermediate-frequency
QPOs only appear in state C of the source, and never appear in state
B. Although it is suggested that the 0.5--2 Hz QPOs may originate
from the compact jet (Fender 2005), the phase-resolved spectra and
variation of the phase lags with frequencies show that the 0.5-10 Hz
QPOs are  all from the inner disk (Miller \& Human, 2005; Reig et
al. 2001). They are linked to the properties of the accretion disk
since their centroid frequencies and fractional rms are correlated
with the thermal flux and the apparent temperature of the disk
(Markwardt, Swank, \& Taam 1999; Trudolyubov et al. 1999; Muno,
Morgan \& Remilland 1999; Sobczak et al. 2000). It is possible that
the QPOs trace the Keplerian motion at the inner radius of the
observable disk (Belloni et al. 2000). We note that the other two
micro-quasars XTE~J1550-564 (Wijnands et al. 1999; Cui et al. 2000)
and GRO~J1655-40 (Remillard et al. 1999; Cui et al. 1999) have
different types of QPOs too. Study of the QPO properties in
GRS~1915+105 can provide important information of the accretion flow
in this source as well as other micro-quasars.

Complex phase/time lags have also been observed for the QPOs of
GRS~1915+105. Cui (1999) found that the hard lags of the
low-frequency QPOs in GRS~1915+105 alternate from negative to
positive values as the frequency increases from the fundamental to
higher harmonic frequencies, and the high frequency QPOs always
show hard phase lag. The intermediate frequency QPOs were
studied by Lin et al. (2000) and Reig et al. (2000). They found
that the hard lags of these QPOs, different from the
phase lags of the low-frequency QPOs, alternate from positive to
negative values as the frequency increases from 0.5 to 10 Hz. The
similar timing characteristics were also observed in XTE~J1550-564
(Cui, Zhang \& Chen 2000). According to the phase lags of
GRS~1915+105, the intermediate frequency QPOs can be further
classified into three types: (1) the 0.5-2 Hz QPOs, whose hard lags
at the fundamental and first harmonic frequencies are both positive;
(2) the 2-4.5 Hz QPOs, whose hard lags are negative at the
fundamental frequency but positive at the first harmonic frequency;
(3) the 4.5-10 Hz QPOs, which show negative phase lags at both the
fundamental and harmonics (Lin et al 2000a; Reig et al. 2000).

It is suggested that the 0.5-10 Hz QPOs observed during state C
provide a link between the optically thick accretion disk and the
Comptonization region (Muno et al. 2001). Since the accretion disk
has temperature structures and the Comptonization region could
up-scatter the lower energy photons into higher energy ones, the
energy dependence of different types of QPOs and their phase lags
can set strong constraints on the current accretion disk models and
provide another avenue to explore the origin of phase lags of
GRS~1915+105. According to the drifting blob model, the centroid
frequency of a QPO is a function of the photon energy. The
energy-dependence of the fractional rms and phase lags of the QPOs
have been widely studied (Morgan et al. 1997; Cui 2000; Cui et al.
1999; Lin et al. 2000a; Reig et al. 2000; Rodriguez et al. 2004).
For $\sim 4$ Hz QPO of XTE~J1550-564, Sriram et al.(2007) found a
frequency difference at two energy bands(2-20keV and 20-50keV).
Choudhry et al. (2005) found that the centroid frequency of the
3$\sim$4 Hz QPOs at 2-7 keV is higher than those in 20-50keV in
GRS~J1915+105. However, as a parameter of the theoretical model of
the QPOs, the relations between the QPO centroid frequency and
photon energy in GRS~1915+105 and other microquasars have not been
well studied yet.

Since photons with different energies are usually from regions with
different physical properties, the energy-dependence of QPOs can
provide additional information that may be critical for a better
understanding of the QPO origins. Therefore, we present in this
paper a study of the energy dependence of the centroid frequencies
of the 0.5-10 Hz QPOs. In section 2, we describe the data and how
they were analyzed. The main results are given in section 3, their
physical implications are discussed in section 4, and section 5 is a
short summary of this work.

\section{Data Reduction and Analyses}

To evaluate the energy-dependence of the QPO centroid frequency, we
select the $RXTE$ observations published in Morgan et al. (1997),
which show the 0.5-10 Hz QPOs and have enough exposure time for each
observation to evaluate the QPOs. In these observations (Table
\ref{table1}), GRS~1915+105 was in class $\chi$ of state C in the
classification by Belloni et al. (2000) or in the plateau state
(Fender 2001), because these observations didn't show strong
variability, HR$_2$(13-60 keV/2-5 keV)$> 0.1$, and the original disk
contribution is expected to be very soft here. The timing and
spectral properties of GRS~1915+105 in those observations have been
widely studied previously. For example, Lin et al. (2000a) and Reig
et al. (2000) have investigated the phase-frequency dependence of
the QPOs, while Muno et al. (2001) and Trudolyubov et al. (1999)
have reported its energy spectral properties. The QPO centroid
frequency is relatively stable over each $RXTE$ epoch in these
observations. Therefore they are very suitable to study the energy
dependence of the centroid frequencies of the QPOs.
% (Lin et al. 2000a; Reig et al. 2000)

\textbf{Table 1}

The data are reduced by using the FTOOLS package as described by Qu,
Yu \& Li (2001). The timing analyses include calculations of the
power density spectrum (PDS) and the cross-power spectrum (CPS),
using the binned mode and event mode data respectively. According to
the data modes of the observations, we extract the light-curves of
GRS~1915+105 with a time resolution of $\sim 4$ ms ($2^{-8}$ s) in
seven PCA energy bands defined in Table \ref{table2}. Among these
energy bands, the hardest energy band (Channels 50-103) has the
lowest statistics, and its mean count rate is still about 320 cts/s
with the model predicted background count rate of 16 cts/s. In every
energy channel, the quality factor ($Q=f/\Delta f_{FWHM}$) of the 6
Hz QPO that presented in the observation is greater than 4,
permitting a detailed study on the relations between the QPO
frequency, photon energy, and phase lag.

\textbf{Table 2}

The PDS is fitted with a model including a power law to represent
the continuum plus one or two Lorentzians to represent the QPOs.
However, it is difficult to obtain a statistically acceptable fit to
the PDS exactly between 1/16 and 16 Hz. For an example, we fit the
1/16 to 16 Hz PDS of observation (10408-01-32-00) that shows a 6 Hz
QPO with a power law plus three Lorentzians, the reduced $\chi^2$ is
6.5. If we limit the frequency range as 4 to 8 Hz, the fit is
improved apparently, with $\chi^2 < 1.6$. Thus in order to get the
accurate centroid frequency and the full width at half maximum
(FWHM) of the QPO in 2-13 keV, the frequency range is selected to
cover the QPO and to make the reduced $\chi^2$ close to the minimum
(see Table \ref{table1}), similar to that used by Cui (1999). The
PDSs in the other energy bands are fitted in the same frequency
range by the model forenamed. The errors of the model parameters are
derived by varying the parameters until $\Delta\chi^2=1$.

%On the other hand, the
%quality of the PDSs at different energy bands are ensured by
%choosing observations (see Table 1) from RXTE data archive and
%suitable energy bands (see Table 2) {\bf (not understood???)}.

The phase lags $\phi$ of the QPOs are calculated by averaging the
phase lags over the frequency range from $f_{\rm QPO}$-${\rm
FWHM}/2$ to $f_{\rm QPO}$+${\rm FWHM}/2$. Their errors are estimated
from the standard deviation of the real and imaginary parts of the
CPSs (Cui et al. 1997). Figure \ref{pds} shows two example PDSs in
energy bands of 2-5 keV and 18-38 keV as well as the CPSs between
these two energy bands. Apparently, the centroid frequency of the
QPO around $\sim$ 6 Hz has a higher value in the harder energy band.
The inset in this figure also shows that the fitting method
above-mentioned gives reasonable fits to the PDSs.

\textbf{Figure 1}

\section{Results}
We find that the centroid frequencies of QPOs are related with
photon energy. This relation evolves from a negative correlation to
a positive one when the QPO frequency increases. Figure \ref{dfreq}a
shows such relations for a few typical QPOs. The energy-dependence
of the centroid frequency of the QPO can be fitted by a power-law,
and the fitted results are listed in Table \ref{table3}. For QPOs
with the centroid frequencies lower than 3 Hz, the centroid
frequency decreases monotonically with photon energy, but the
correlation becomes weaker with the centroid frequency increases.
For QPOs with the centroid frequencies higher than 3 Hz, the
centroid frequency increases significantly with photon energy, and
the correlation also becomes stronger as shown by the correlation
coefficients. However, for QPOs around 3 Hz, their centroid
frequencies don't have a monotonic evolutionary trend with photon
energy, while the values of the correlation coefficients turn over
their sign from negative to positive.

\textbf{table 3}

The relation between phase lag and photon energy is opposite to that
between QPO centroid frequency and photon energy. The results of
fitting and correlation coefficient are listed also in Table
\ref{table3} and displayed in Figure \ref{plags}b. When the QPO
frequency is around 1 Hz, the phase lag is positively correlated
with photon energy, and the two quantities become negatively
correlated when the QPO frequency reaches above 3.5 Hz. These
results are similar to the ones of Lin et al. (2000a) and Reig et
al. (2000).

\textbf{Figure 2}

In Figure \ref{qpolags} we plot the centroid frequency variation
$\Delta f$ and phase lag $\phi$ versus the QPO frequency for all the
QPOs we detected in the observations. Both $\Delta f$ and phase lag
are calculated between two energy bands of 2-5 keV and 13-18 keV
only. It is shown that {$\Delta f$} increases with the QPO
frequency, while the phase lag decreases, indicating a negative
correlation between $\Delta f$ and $\phi$ for the QPOs.

\textbf{Figure 3}

The negative correlation between $\Delta f$ and $\phi$ holds not
only among different QPOs but also within a QPO.  The $\Delta f$ and
$\phi$ calculated between 2-5 keV and the other five higher energy
channels (Table 2) for a few typical QPOs are shown in Figure
\ref{philag}.  The results show that $\Delta f$ and $\phi$ has an
anti-correlation for QPOs with centroid frequency lower than 2 Hz or
higher than 3.5 Hz, and no correlation for QPOs between 2 and 3.5
Hz. We calculate the correlation coefficients and fit the relation
with a linear function to all the QPOs we detected in this work. The
results are listed in Table \ref{table4}.

\textbf{Table 4}

\section{Discussions}
For the first time we find the centroid frequency evolution with
photon energy for the 0.5-10 Hz QPOs of GRS~1915+105. We also find
that the QPO phase lag is correlated with both photon energy and QPO
frequency. The QPO centroid frequencies are shown to have an
anti-correlation with the phase lags, as shown in Figure
\ref{philag}. These results set strong constraints on the current
models of the origin of QPOs and phase lags in black hole binaries,
and make a direct challenge for theorists to explain the new
observable phenomena.

Various models have been proposed to explain the timing phenomena in
black hole binaries. It is generally believed that the X-ray
radiation of a black hole binary is contributed by three components:
the soft X-ray radiation from the accretion disk, the hard
components from the Compton cloud and/or jet (McClintock \&
Remillard 2003). The QPOs are suggested to be related to the
accretion disk of the compact object, while the phase lags to the
electron cloud. Particularly, the 0.5-10 Hz QPOs of GRS~1915+105
could occur in the inner region of the disk and are associated with
disk instabilities (see van der Klis 2004 and McClintock \&
Remillard 2003 for review). However, GRS~1915+105 displays the
number of X-ray states, many models for microquasar behavior are
based on a limited number of its X-ray states. Based on our
observational results, we only discuss the following four models
commonly used to to describe the origin of QPOs in compact objects:
(1) The global disk oscillation (GDO) model; (2)The radial and
orbital oscillation model (ROOM); (3) The accretion flow instability
model (AFIM); and (4) the drift blob model (DBM).

\subsection{Constraints from the energy dependent QPO frequency}
For orbital and epicyclic frequency models, the particles moving
around the compact objects could have different oscillating
frequencies in the inner region of the accretion disk: orbital,
radial and vertical epicyclic frequencies. Damping, or the
superposition of many local frequencies can turn the intrinsically
periodic disk oscillations into QPOs or broad noise. In the global
disk oscillation (GDO) model, the disk oscillation is a vertical
mode, and the GDO frequency is expected to be also independent of
the photon energy and should be seen in all the energy bands that
disk emits (Titarchuk 2000). Thus the GDO model can not explain
the observational phenomena of the intermediate-frequency QPOs.

For the radial and orbital oscillation model (ROOM), the oscillation
frequency is a function of the disk radius/temperature (Nowak \&
Wagoner 1993; Nowak 1994). The QPO frequency will vary with energy
because photons with different energies are from different radii.
These models can therefore explain qualitatively the evolution of
the QPO frequency with energy. However, the emission from the disk
is thermal. There should be a break of the QPO power represented by
the root-mean-square (RMS) at higher energy if the oscillations
occur in the inner region of the disk, since the high energy
emission is from the Compton cloud (Rodriguez et al. 2004). However,
such a break has not been detected up to $\sim 30$ keV, making the
disk oscillation model of the QPOs unlikely. Meanwhile, the ROOM can
not explain the various correlations between the centroid frequency
and photon energy for different QPOs. The ROOM model need to be
greatly modified in order to fit the observational results.

In the accretion flow instability model (AFIM), locally at each
radius the disk fluctuates on different instability timescales, and
the oscillations propagate in the disk (Nowak 1994). In this
scenario, emission from the inner region of the disk tends to have a
higher QPO frequency and a harder spectrum. This model can naturally
explain the observed positive correlation between the centroid
frequency and photon energy for the QPOs with frequency higher than
$\sim$3.5 Hz. But the observed negative correlation for QPOs less
than 2 Hz contradicts the model prediction.

The energy-dependence of the QPO frequency is expected by the drift
blob model (DBM; B\"ottcher \& Liang 1998, 1999; Hua et al. 1997).
In this model, the Keplerian motion of the blobs could manifest
itself in a QPO observationally, and any radial drift of the blobs
would cause the QPO frequency to increase with energy. Similar to
the AFIM model, the DBM model can explain the observed properties of
the QPOs higher than $\sim$3 Hz, but not for the QPOs less than 2
Hz. The relation between model and observed phenomena is summarized
in table \ref{tablemodel}.

Of the models we considered above, none can fully explain the energy
dependencies of QPOs obtained in this paper. The energy dependencies
of the QPO centroid frequencies provide additional information for
theoretical models. And the new observable properties of the QPOs of
the GRS~1915+105 also make a challenge to the new models of the
QPOs, i.e, the new theoretical model should not only reproduce or
explain the observable phenomena in this paper, but also explain the
other properties such as the energy dependence of the rms and the
phase lags (Cui 1999, Rodriguez et al. 2004).

\textbf{Here insert Table 5}

\subsection{Phase lag results and their constraints on models}

To explain the observational phase (time) lags in the compact X-ray
objects, a lot of models are proposed (see Cui 1999; Poutanen 2001
for review, and reference therein). Among them, the so called
standard model and the perturbation propagation model are the two
major ones. In the standard model, the time lag is considered as the
diffusion timescale of the photon passing through the Comptonization
region (Cui 1999; Poutanen 2001). The hard lag is the result of the
process in which the soft (seed) photons gain energy from the hotter
electron corona. For the perturbation propagation model, the soft
phase lags can be explained by assuming that perturbations propagate
from the inner disk to the outer disk. There may also be
perturbations propagating inward from the outer edge of the disk,
which generate hard phase lags (Lin et al. 2000b).

The observed phase lag behaviors in this paper challenge the above
two models.  Since the standard model only predicts hard lags, the
measured soft lags in GRS~1915+105 are incompatible with the model
(see also Cui 1999), and furthermore, the observed large time lag
values require a huge corona size ($\sim 3\times 10^{10}$cm) that is
physically unrealistic (Hua, Kazanas \& Cui 1999). Although the
perturbation propagation model can explain the observed soft and
hard lags qualitatively, it is not clear how the propagation
direction of the perturbation transits when QPO frequency passes 2
Hz, at which the phase lag of the QPO changes sign as revealed in
this work. Probably the phase lags are of multiple origins.

The anti-correlations between the phase lags $\phi$ and frequency
variation ($\Delta f$) of the QPOs (see also Figure \ref{philag}),
and between $\Delta f$ and $\log(E)$, imply that the phase lags
could be caused by the change of the QPO frequency, which may give a
new approach to explain the phase lags in the compact X-ray objects.
To better illustrate this point, we derived the relation between the
phase lag and $\Delta f$ as follows.

 According
to the definition of the cross correlation function,
$ccf(\tau)=\int_{-\infty}^{\infty}h(t)s(t-\tau)dt$, the cross power
spectrum is $CPS(\nu)=\int_{-\infty}^{\infty}ccf(\tau)e^{i2\pi
f\nu\tau}d\tau=S^*(\nu)H(\nu)$, i.e.,
\begin{eqnarray*}
CPS(\nu)&=&|S^*|e^{-i\phi_s(\nu)}|H|e^{i\phi_h(\nu)}\\
&=&|S^*H|e^{i(\phi_h(\nu)-\phi_s(\nu))}\\
&=&Ae^{i\phi(\nu)}
\end{eqnarray*}
where, $S^*(\nu)=|S^*|e^{-i\phi_s(\nu)}$ and
$H(\nu)=|H|e^{i\phi_h(\nu)}$. If the light curves with oscillations
in two different energy bands could be described by $s(t)=s(f t)$
and $h(t)=h((f+\Delta f)t)$, then,
\begin{eqnarray*}
CPS(\nu) &=& \frac{1}{f(f+\Delta f)}S^*(\frac{\nu}{f})H(\frac{\nu}{f+\Delta f})\\
&=&Ae^{i[\phi_h(\frac{\nu}{f+\Delta f})-\phi_s(\frac{\nu}{f})]}\\
\end{eqnarray*}
So the phase lag is,
\begin{eqnarray*}
\phi(\nu) &=&
\phi_h(\frac{\nu}{f+\Delta f})-\phi_s(\frac{\nu}{f})\\
&\approx&
\phi_h(\frac{\nu}{f}(1-\frac{\Delta f}{f}))-\phi_s(\frac{\nu}{f})\\
&\approx&
\phi_h(\frac{\nu}{f})+\phi_h^\prime|_\frac{\nu}{f}\times\frac{\nu\Delta
f}{f^2}-\phi_s(\frac{\nu}{f})
\end{eqnarray*}
If let $\nu=f$, the phase lag of the QPO with frequency $f$ is,

\begin{eqnarray*}
\phi(f) &=& \phi_h(f)-\phi_s(f)+\phi_h^\prime(f)\times\frac{\Delta
f}{f}\\
&=&\phi_{\lambda}(f)+\phi_f(\Delta f)
\end{eqnarray*}
where, $\phi_{\lambda}(f)=\phi_h(f)-\phi_s(f)$ is phase lag due to
physical processes such as Comptonization and  $\phi_f(\Delta
f)=\phi_h^\prime(f)\times\frac{\Delta f}{f}$ is caused by the change
of the oscillation frequency of the QPO. $\phi_f(\Delta
f)=k_{model}\Delta f=k_f\frac{\phi_h^\prime}{2\pi}\Delta f$
($k_{model}\propto k_f=\frac{2\pi}{f}$), i.e., the phase lag caused
by the change of the QPO frequency is proportional to $-\Delta f$.

If the observed phase lag is due to the change of the QPO frequency,
the $k_{obs}$ should be proportional to $k_{f}$, where $k_{obs}$ is
the fitted slop between $\phi$ and $\Delta f$. In Table \ref{table4}
we list $k_f$ and $k_{obs}$ as well as the correlation coefficient
between them. However, there is not an obvious correlation between
$k_{obs}$ and $k_{f}$, which means that the measured phase lags are
caused partially by the change of the QPO frequency. These results,
together with $k_{obs}$ and $k_{f}$, show that the phase lag caused
by $\Delta f$ is a small fraction of the observed phase lags for
QPOs less than 2 Hz but becomes dominant for QPOs with $f\geq$3.5
Hz.

\section{Summary}
We find that the frequency and phase-lag are both energy dependent
for the 0.5-10 Hz QPOs in GRS~1915+105. For QPOs with the centroid
frequencies less than 3 Hz, the centroid frequency decreases
monotonically with photon energy. For QPOs with the centroid
frequencies larger than 3 Hz, the centroid frequency increases
significantly with photon energy. However, for the QPOs around 3 Hz,
the centroid frequencies of the QPOs don't have a monotonic
evolutionary trend with photon energy (see Figure 2 and Table 3).
Meanwhile, the phase lag behaviors are closely related to the
variables of the QPO centroid frequency. The phase lag is negatively
correlated with the centroid frequency difference of the QPO when
QPO frequency larger than 3.5 Hz or less than 2 Hz. No correlation
for phase lag and centroid frequency of the QPOs is found between 2
and 3.5 Hz. Of the models we considered, none can fully explain all
the properties of QPOs obtained in this paper (see Table
\ref{tablemodel}).

It is shown that the observed phase lag may be caused by two
mechanisms, one is the comptonization of the soft photon, another is
caused by the frequency difference of the QPO at different energy
bands. The frequency difference between two energy bands could be a
main source of the phase-lag for QPOs with $f\geq$ 3.5Hz. However,
it is necessary to further investigate the energy dependence of the
QPOs in micro-quasars for verification generally. Meanwhile, many of
the models proposed are actually variability-frequency models, which
predict the frequency, or the power spectrum, of the fluctuations
only in physical flow parameters rather than in any observable
quantity. In order to allow further tests of QPO models and to
discriminate between them, predictions of the observable of the
oscillations are essential. However, the new phenomena in this paper
make a direct challenge to the theoretical model for the QPO. A
successful model should be able to reproduce the energy-dependencies
of the QPO centroid frequencies we observed, but also the
correlations between $\Delta f$ and phase lags should be explained.

\acknowledgements{The authors would like to thank anonymous referee
for some helpful suggestions and comments. This work was partially
supported by the National Basic Research Program of China (Grant No.
2009CB824800), the Natural Science Foundation of China (Grant No.
10773017), the Natural Science Foundation of Xinjiang Uygur
Autonomous Region of China (Grant No. 200821164), and the Program of
the Light in Chinese Western Region (LCWR)(Grant No. LHXZ 200802)

%\begin{thebibliography}{}

\begin{figure}
\plotone{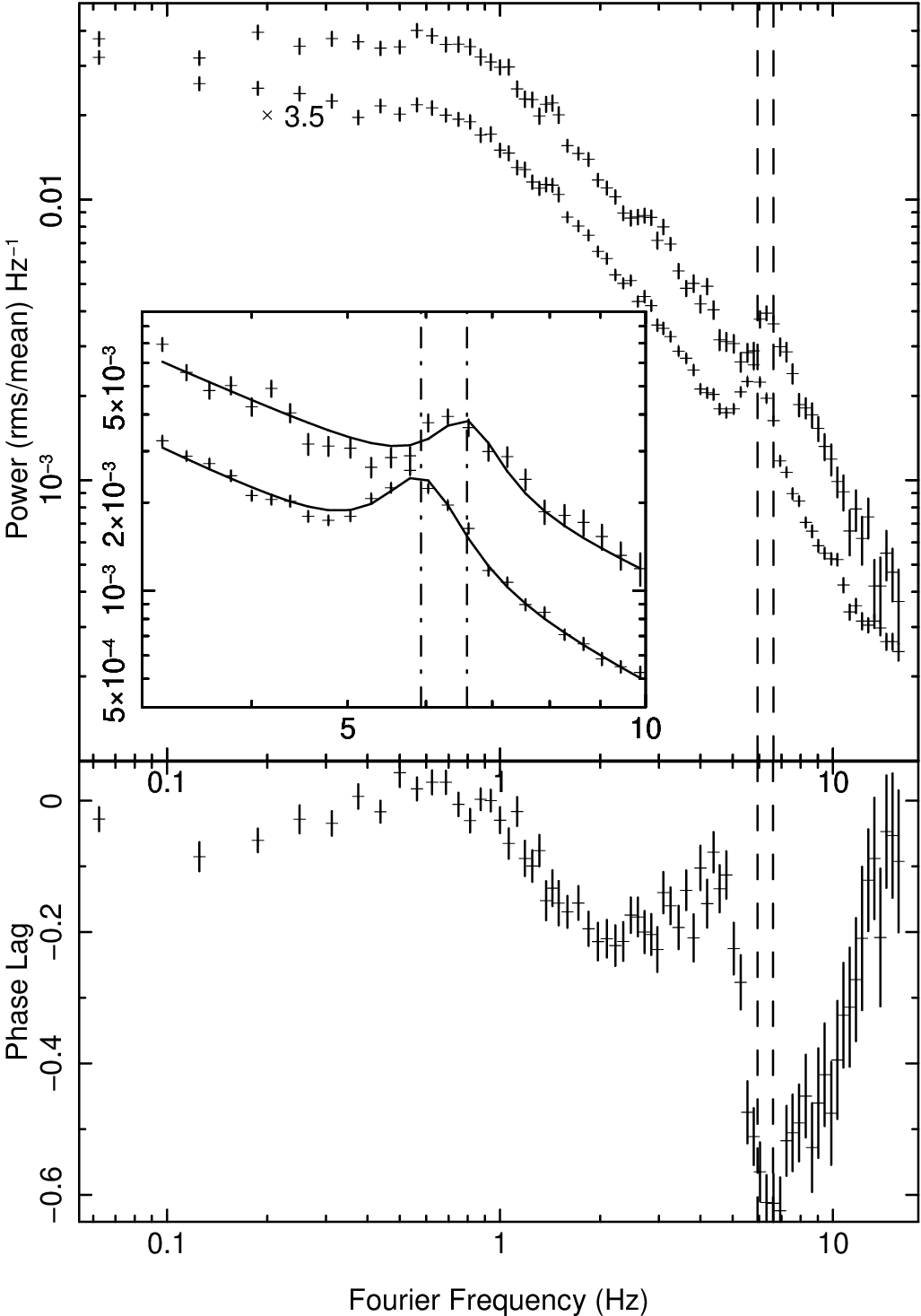} \caption{Power density spectrum (upper) and cross
power spectrum (bottom) of GRS~1915+105 in observation
10408-01-32-00. The inset shows the frequency difference of the $~6$
Hz QPO in energy bands 2-5 keV and 13-18 keV.\label{pds}}
\end{figure}

\begin{figure}
\plotone{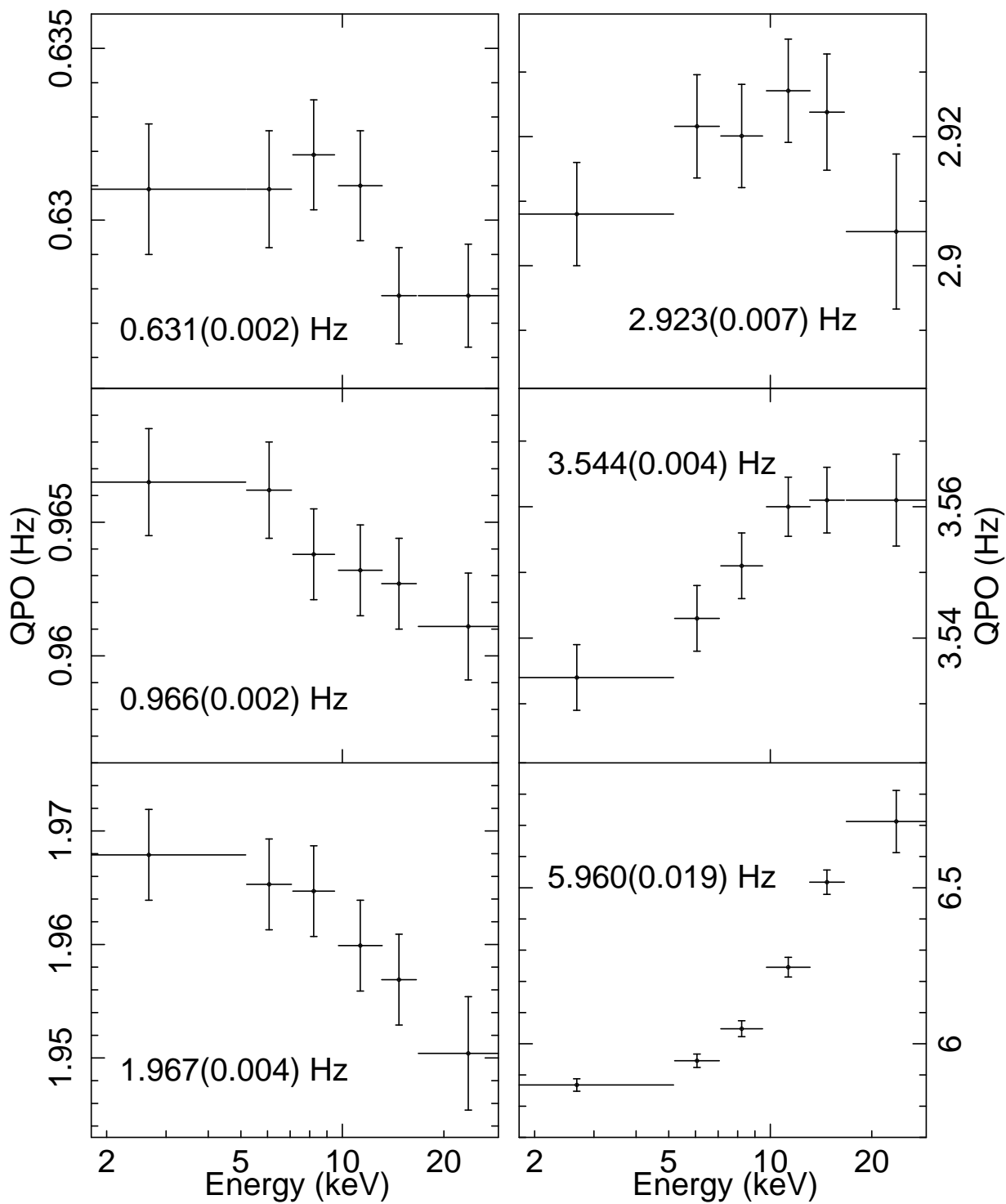} \caption{a): The relation between the QPO
centroid frequency and photon energy for a few typical QPOs. The
digit in each panel is the centroid frequency of the QPO at
2-13~keV. \label{dfreq}}
\end{figure}

\addtocounter{figure}{-1}
\begin{figure}
\plotone{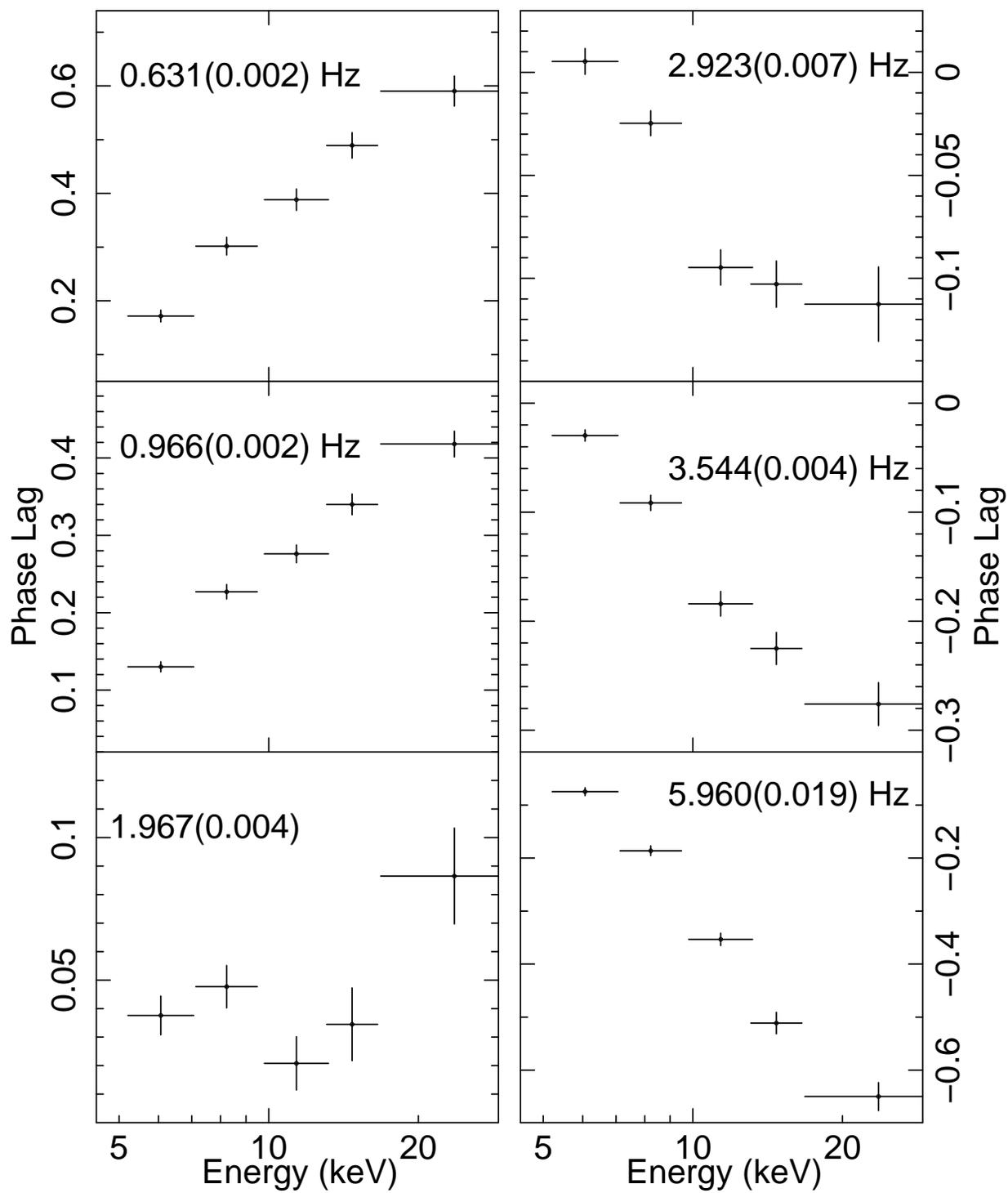} \caption{b): The relation between phase lag and
photon energy for a few typical QPOs.\label{plags}}
\end{figure}

\begin{figure}
\plotone{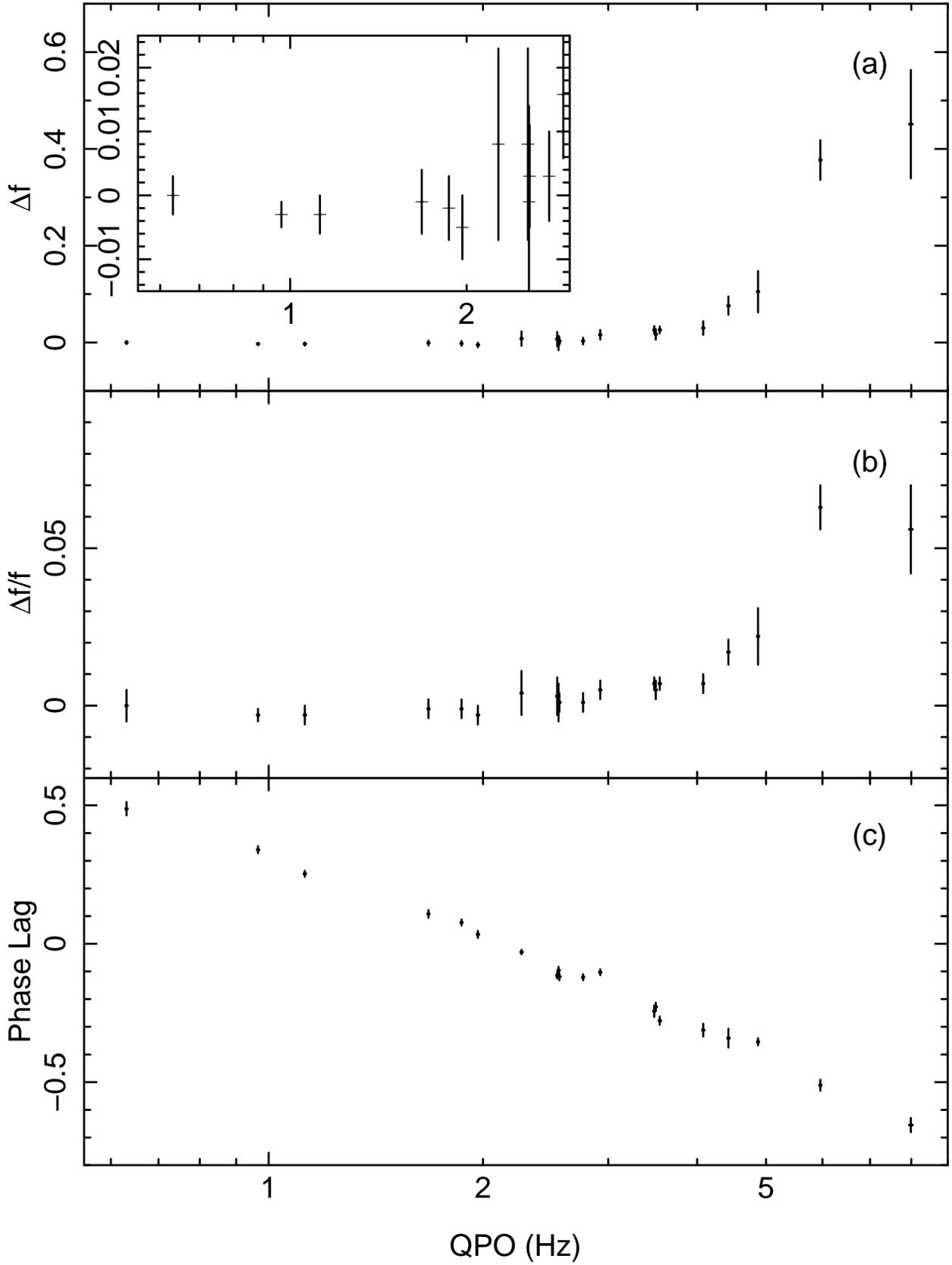}%{QPO_Phase.ps}%
\caption{QPO frequency differences and phase lag vs QPO frequency
for energy bands 2-5 keV and 13-18 keV. The inset show the $\Delta
f$ of the QPOs with frequencies less than 3 Hz. \label{qpolags}}
\end{figure}

\begin{figure}
\plotone{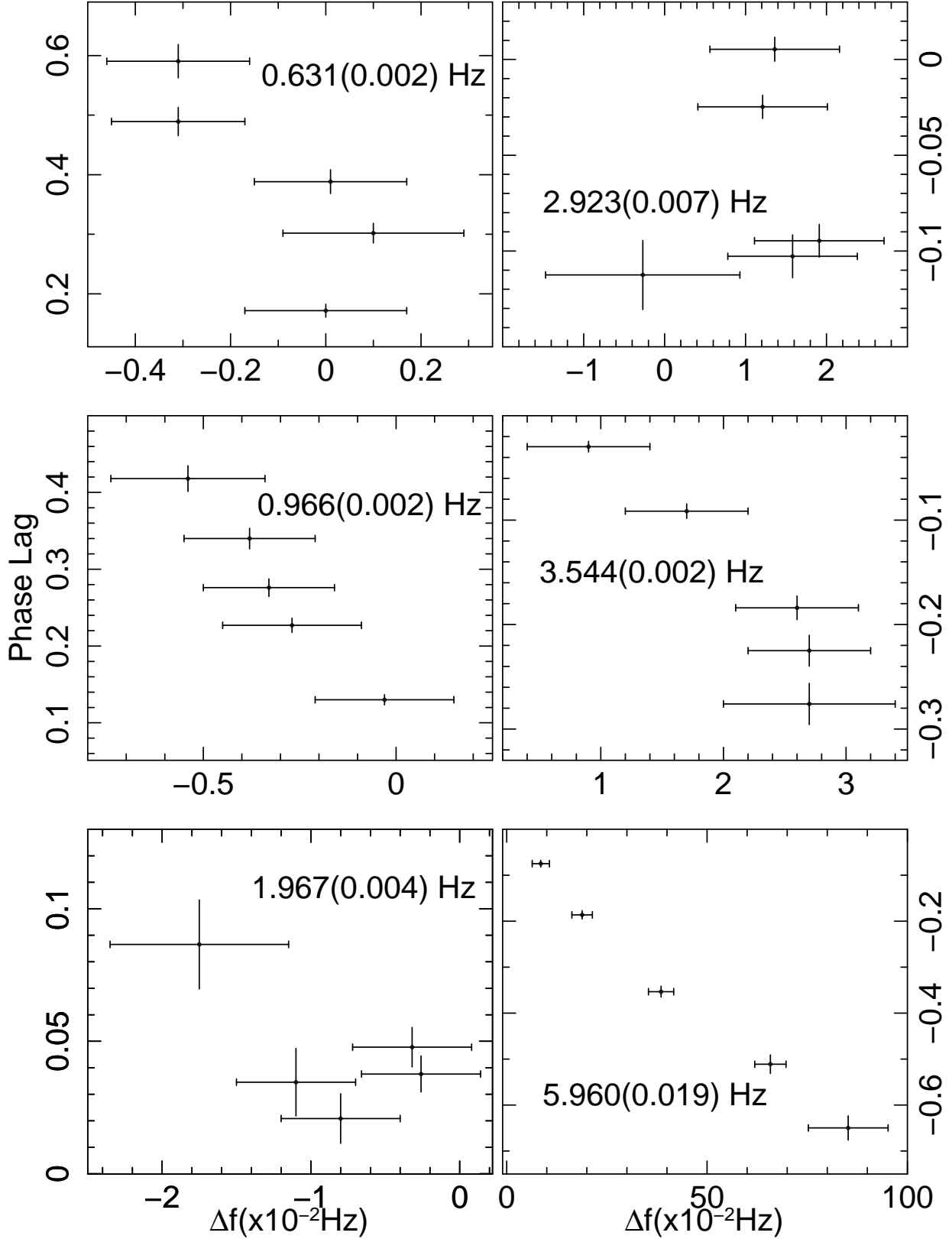}%{DelQPO_Phase.vps}%
\caption{Frequency differences($\Delta f$) vs phase lags ($\phi$)
for various typical QPOs. The digits are the centroid frequencies of
the QPOs at 2-13 keV. Refer to the text for detail. \label{philag}}
\end{figure}

\begin{deluxetable}{ccccc}
\tablecolumns{5} \tablewidth{0pc} \tablecaption{The RXTE
Observations of GRS~1915+105 used in this paper} \tablehead{ ObsID &
Date  &  Exposure(s)  & $f_c$/(Fitting range)} \startdata
21-02  & 07/07/96 & 2000 & $7.988\pm0.051$/(4.5-12)  \\
22-00  & 11/07/96 & 2520 & $3.479\pm0.005$/(1.8-6.5) \\
22-01  & 11/07/96 & 3320 & $2.770\pm0.005$/(1.7-9)   \\
22-02  & 11/07/96 & 3312 & $2.554\pm0.006$/(0.16-4.2)\\
23-00ab&14/07/96 & 6640 & $3.544\pm0.005$/(2-10)     \\
24-00ab  &16/07/96 & 6544 & $2.266\pm0.006$/(1.6-6.5)\\
24-00cd  &16/07/96 & 6544 & $2.543\pm0.005$/(1.6-6.5)\\
25-00   &19/07/96 & 3328 & $1.125\pm0.002$/(0.2-1.8) \\
27-00   &26/07/96 & 8960 & $0.631\pm0.002$/(0.4-2)   \\
28-00   &03/08/96 & 9984 & $0.966\pm0.002$/(0.2-2.9) \\
29-00a  &10/08/96 & 2952 & $1.677\pm0.004$/(0.8-4.6) \\
29-00b  &10/08/96 & 3392 & $1.866\pm0.004$/(0.8-4.6) \\
29-00c  &10/08/96 & 3392 & $1.967\pm0.004$/(0.8-4.6) \\
30-00   &99/08/98 & 9960 & $4.871\pm0.011$/(1.8-10)  \\
31-00a  &25/08/96 & 2208 & $4.092\pm0.007$/(2.6-6)   \\
31-00b  &25/08/96 & 2912 & $4.434\pm0.011$/(2.5-8.8) \\
31-00c  &25/08/96 & 2912 & $3.505\pm0.008$/(2.2-7.4) \\
32-00   &31/08/96 & 7424 & $5.960\pm0.019$/(3-10)    \\
**49-00 &08/10/97 & 3408 & $2.923\pm0.007$/(1.4-15)\\
\enddata\label{table1}
21-02=10408-01-21-02; ab=a+b; *49-00=20402-01-49-00;

$f_c$ is the QPO frequency at PCA channel 2-13 keV
\end{deluxetable}

\begin{deluxetable}{ccc}
\tablecolumns{3} \tablewidth{0pc} \tablecaption{RXTE/PCA Energy to
Channel Table} \tablehead{PCA Channel  &   Energy Range(keV)   &
Centroid Energy (keV) } \startdata
0-35   &  1.94-12.99  & $ 6.39$\\
0-13   &  1.94-5.12   & $ 2.67$\\
14-18  &  5.12-6.89   & $ 6.06$\\
19-25  &  6.89-9.39   & $ 8.22$\\
26-35  &  9.39-12.99  & $ 11.30$\\
36-49  &  12.99-18.09 & $ 14.72$\\
50-103 & 18.09-38.44  & $ 23.63$\\
\enddata\label{table2}
\end{deluxetable}

\begin{deluxetable}{ccccccccc}
\tablecolumns{9} \tablewidth{0pc} \tablecaption{Energy-dependence of
the QPO frequencies and phase lags} \tablehead{ \colhead{}
&\colhead{} & \multicolumn{3}{c}{E$-f_{\rm QPO}$} & \colhead{}  &
\multicolumn{3}{c}{E$-\phi$}    \\
\cline{3-5} \cline{7-9} \\
\colhead{ObsID} & \colhead{$f_c$} &
\colhead{$\Gamma$($\times10^{-3}$)}&\colhead{$\chi^2$} &
\colhead{Cor}& \colhead{$\phi^+$} & \colhead{$\Gamma$} &
\colhead{$\chi^2$}&\colhead{Cor} } \startdata
21-02 & $7.988\pm0.051$ & $45.0\pm6.7$ & 6.1  & 0.98 &-$0.655\pm0.026$ & $1.17\pm0.33$ &1.75 &-0.97\\
32-00 & $5.960\pm0.019$ & $49.0\pm3.1$ & 76.5 & 0.97 &-$0.511\pm0.020$ & $0.97\pm0.13$ &5.98 &-0.96\\
31-00a& $4.092\pm0.007$ & $3.1\pm1.3 $ & 3.2  & 0.78 &-$0.312\pm0.023$ & $0.69\pm0.39$ &1.01 &-0.92\\
23-00 & $3.544\pm0.005$ & $4.3\pm0.9 $ & 2.1  & 0.81 &-$0.278\pm0.015$ & $0.78\pm0.42$ &0.75 &-0.95\\
22-00 & $3.479\pm0.005$ & $3.9\pm1.0 $ & 2.7  & 0.83 &-$0.244\pm0.021$ & $0.80\pm0.50$ &0.35 &-0.75\\
*49-00& $2.923\pm0.007$ & $-1.2\pm1.8$ & 3.8  &-0.29 &-$0.103\pm0.011$ & $0.45\pm0.35$ &0.06 &-0.87\\
22-01 & $2.770\pm0.005$ & $-1.1\pm1.4$ & 6.4  &-0.71 &-$0.121\pm0.011$ & $0.61\pm1.6 $ &0.08 &-0.91\\
22-02 & $2.554\pm0.006$ & $-6.6\pm1.4$ & 3.4  &-0.69 &-$0.096\pm0.013$ & $0.63\pm5.0 $ &0.02 &-0.92\\
29-00c& $1.967\pm0.004$ & $-3.2\pm1.4$ & 1.1  &-0.99 &$0.034\pm0.013 $ & $0.62^{+0}_{-\infty}$&0.01&0.57\\
25-00 & $1.125\pm0.002$ & $-4.2\pm1.3$ & 1.94 &-0.997&$0.253\pm0.011 $ & $0.65\pm0.26$ &0.453& 0.93\\
28-00 & $0.966\pm0.002$ & $-3.2\pm0.2$ & 0.48 &-0.94 &$0.340\pm0.013 $ & $0.66\pm0.15$ &1.35 &0.94 \\
27-00 & $0.631\pm0.002$ & $0.52\pm1.5$ &2.58  &-0.80 &$0.489\pm0.024 $ & $0.067\pm0.027$&2.0 &0.94 \\

\enddata\label{table3}
Cor=Correlation coefficient; $\Gamma$ is power index of the relation
between $\Delta f$ or $\phi$ and photon energy $E$.

The phase lag $\phi^+$ is obtained between 13-18 keV and 2-5 keV
energy bands; Reduced~$\chi^2$=$\chi^2/(6-2)$
\end{deluxetable}

\begin{deluxetable}{cccccc}
\tablecolumns{6} \tablewidth{0pc} \tablecaption{The relation between
frequency differences ($\Delta f$) and phase lags ($\phi$) of the
QPOs} \tablehead{ \colhead{ObsID} & \colhead{$f_c$}
&\colhead{$k_{obs}$} &\colhead{$\chi^2$}&\colhead{Cor}
&\colhead{$k_{f}$} } \startdata
21-02 & $7.988\pm0.051$ & $-1.08\pm0.30$   &1.47  &-0.99  &-0.79\\
32-00 & $5.960\pm0.019$ & $-0.694\pm0.093$ &0.677 &-0.99   &-1.1\\
31-00a& $4.092\pm0.007$ & $-1.1\pm7.4$     &4.15  &-0.90  &-1.5\\
23-00 & $3.544\pm0.005$ & $-12.9\pm6.8$    &0.7   &-0.94  &-1.8\\
22-00 & $3.479\pm0.005$ & $-8.0\pm8.0$     &2.0   &-0.59  &-1.8\\
49-00& $2.923\pm0.007$  & $0.5\pm8.4$      &0.10  &-0.08   &2.1\\
22-01 & $2.770\pm0.005$ & $2.3\pm6.4$      &0.15  &0.48   &2.3\\
22-02 & $2.554\pm0.006$ & $3.5\pm17 $      &0.03  &0.64   &2.5\\
29-00c& $1.967\pm0.004$ & $-3.3\pm18$      &0.01  &-0.78  &-3.2\\
25-00 & $1.125\pm0.002$ & $-20.3\pm8.5$    &1.38  &-0.90  &-5.6\\
28-00 & $0.966\pm0.002$ & $-59\pm13$       &0.86  &-0.97  &-6.5\\
27-00 & $0.631\pm0.002$ & $-65\pm15$       &9.1   &-0.81  &-10\\

\enddata\label{table4}
$k_{f}=2\pi/f_{\rm QPO}$; $k_{obs}$ is the fitted slope of the
relation between $\phi$ and $\Delta f$.

Reduced~$\chi^2$=$\chi^2/(5-2)$
\end{deluxetable}
separate

\begin{deluxetable}{rccl}
\tablecolumns{4} \tablewidth{0pc} \tablecaption{Model and
Observation} \tablehead{ Model & Prediction(E$\sim\Delta f$) &
Observation  & Reference} \startdata
  GDO    & $\Delta f=0$         & no     & Titarchuk 2000\\
  ROOMs  & $\Delta f\sim f(E)$  &partial & Nowak \& Wagoner 1993; Nowak 1994\\
  AFIMs  & $\Delta f\sim f(E)$  &partial & Nowak 1994\\
  DBM    & $\Delta f\sim f(E)$  &partial & B\"ottcher \& Liang 1998, 1999; Hua et al. 1997 \\
\enddata\label{tablemodel}
GDO=the global disk oscillation\\
ROOMs=the radial and orbital oscillation models\\
AFIMs=the accretion flow instability models\\
DBM=the drift blob model\\
partial=The model only explain partial observed phenomena.
\end{deluxetable}

\end{document}